# MAGAL Constellation – Using a Small Satellite Altimeter Constellation to Monitor Local and Regional Ocean and Inland Water Variations

André G. C. Guerra[1], André João[1], Miguel Arantes[1], Miguel Martin[1], Paulo Figueiredo[1], Alexander Costa[1],[6], Catarina M. Cecilio[2], Inês Castelão[2], Clara Lázaro[3], Joana Fernandes[3], A. Marques[4], K. Brandão[5], P. Lima[5], Yaroslav Mashtakov[6], Anna Guerman[6], Catharina Pieper[7], Ana Martins[7], Burke O. Fort[8], Timothy J. Urban[8], Byron D. Tapley[8], Brandon A. Jones[9]

[1] CEiiA, Av. Dom Afonso Henriques, 1825. 4450-017 Matosinhos, Portugal, (+351) 220 164 800, space@ceiia.com

[2] CoLAB +ATLANTIC, Edifício LACS, Estrada da Malveira da Serra 920, 2750-834 Cascais, Portugal, info@colabatlantic.com

[3] FCUP/CIIMAR, Rua do Campo Alegre, 687, 4169-007 Porto, (+351) 220 402 489/Terminal de Cruzeiros de Leixões, 4450-208 Matosinhos, clazaro@fc.up.pt, mjfernan@fc.up.pt

[4] EFACEC, Via de Francisco Sá Carneiro, Apartado 3078, 4471-907 Moreira da Maia, Portugal, (+351) 229 402 000, aerospace@efacec.com

[5] Omnidea, Polypark - Núcleo Empresarial da Arruda dos Vinhos, Bloco A, Piso 0 Escritório 1, Estrada da Quinta de Matos nº 4, 2630-179, (+351) 211 913 169, info@omnidea.net

[6] Universidade da Beira Interior, Rua Marquês D'Ávila e Bolama, 6201-001 Covilhã (+351) 275 329 915, yarmashtakov@gmail.com

[7] OKEANOS - Institute for Research in Marine Sciences, Department of Oceanography and Fisheries, Faculty of Sciences and Technology, University of the Azores, Rua Prof. Dr. Frederico Machado, 4, Horta, Azores 9901-862, Portugal, catharina.pieper@gmail.com

[8] Center for Space Research, The University of Texas at Austin, 3925 West Braker Lane, Suite 200, Austin, TX 78759, USA, (+1) 512 471 0967, fort@csr.utexas.edu

[9] Department of Aerospace Engineering & Engineering Mechanics, The University of Texas at Austin, Code C0600, Austin, Texas USA 78712, (+1) 512 471 4743, brandon.jones@utexas.edu

## ABSTRACT

MAGAL lays the foundations for a future constellation of small satellites carrying radar altimeters aiming to improve the understanding of ocean circulation variability at local, regional, and global scales. All necessary tools will be developed, including a new small, low-power altimeter payload and a miniaturized satellite platform, grounded on the Space 4.0 industry, to be manufactured in-series, minimizing production, operational and launch costs. To implement a collaborative constellation, and better tackle the gaps of large radar altimeter programmes, MAGAL will use a Data Analysis Centre, based on cloud services, for storage and process of data, based on known and improved algorithms, including overlay of layers from multiple sources (e.g. meteorology and open-source data). As a constellation of six satellites, MAGAL increases the density of sea surface topography measurements, enabling more data for altimetry products, when used in synergy with other missions, in coastal areas and over mesoscale features. This results in scientific and commercial information aggregated into a single platform, displayed in various graphical interfaces, allowing overlaid correlations. MAGAL is aligned with the insights from the EU agenda for sustainable development, adding value, alongside the underlying technology development, bringing together the sea's economy and its sustainable growth.

# 1 INTRODUCTION

Satellite altimetry has substantially advanced the understanding of ocean circulation by providing unprecedented observations of sea surface topography, improving the knowledge from the role of mesoscale eddies to global ocean circulation and sea level rise [1, 2, 3]. These measurements have been secured by large conventional spacecraft, which have assured continuity of the climate record for many years. However, there is a need to observe ocean dynamics at smaller (from 15 to 200 km) and faster (shorter than 5 days) scales. Understanding how small-scale dynamics near the coast are linked to the open ocean and how those dynamics impact ocean variability at global and regional scales is still one of the challenges for modern oceanography [4].

Considering the high cost and the main limitations of current traditional altimetry missions (e.g. spatial and temporal coverages determined by the satellite orbit; and the fact that measurements are collected along-track only), extending the current ocean space observational capability by adding more conventional missions remains complicated. In this context, new resources such as a small radar altimeter operating in a constellation, can provide complementary capabilities. Constellations of radar altimetry satellites are, in fact, an emerging concept [5] that can generate more sea surface topography measurements, be a cost-effective alternative, and a complement to the large missions, providing a greater return on investment [6].

The development of constellations of small satellites, in particular micro and nano satellites (i.e. less than 100 kg and less than 10 kg, respectively), has recently grown considerably. Chiefly as a result of the increasing on-orbit capability and reduced system cost [7], allowing a push for democratized access to space, performing increasingly difficult tasks in smaller form factors and at lower costs. The popularity of the CubeSat standard and the availability of low-cost, high-performance components also contribute to the use of new technology. The space sector has thus changed its profile, entering the so called *NewSpace*, which is characterized by a miniaturization tendency and the use of components which were not originally developed for space.

The short-to-medium term planning of altimetry missions outlines the preservation of the ongoing long-term record of reference missions and ensures that new complementary missions are supported by the reference background that has been established. Consequently, by minimizing discontinuity and providing data that can be used in synergy with past records, small satellite constellations can fulfil the stringent requirements of observing global sea level and lead to a better multi-mission altimetric dataset, in particular for local and regional scales.

MAGAL is a future constellation of small satellites carrying radar altimeters that has a foundation based upon the most general recommendations by the International Altimetry Team [1] for the future of altimetry. The measurements provided by MAGAL are, thus, expected to sustain and to improve existing observations from well-established successful altimetry missions and to better observe the global open ocean at finer space and time scales, while also extending observations into coastal and high-latitude oceans.

This paper discusses, in Section 2, the use cases developed in the scope of the mission and describes the development of mission characteristics, payloads and the Data Analysis Centre (DAC) to process the acquired data and combine it with the existing one (Section 3). Section 4 summarizes the mission aspects and the next steps.

## 2 USE CASES

Four main use cases have been selected assuming that the MAGAL constellation will acquire measurements of sufficient accuracy and that the satellite and constellation characteristics are selected to improve the spatial and temporal sampling of the ocean, compared to those of the data acquired by current operational satellite altimetry missions. The selected cases, described below, are expected to improve the knowledge of sea level on local, regional, and global scales, both over ocean and coastal zones. This mission shall also allow a better understanding of ocean processes that occur on meso and sub-mesoscales, which remain poorly understood due to the lack of adequate *in situ* sampling, but also because of limited past and current satellite altimetry missions [8], at diverse ocean and inland water bodies levels.

### 2.1 Sea Surface Topography

The minimum requirement for mesoscale characterization is the simultaneous use of two satellite altimetry missions, provided one is a reference mission (e.g. Jason-3 or Sentinel-6 Michael Freilich) and the other a complimentary one (e.g. Sentinel-3) [8]. However, studies have shown that when combining data from three or more missions, the root mean square (RMS) of sea level anomaly and eddy kinetic energy differences are significantly reduced [9], leading to an improved characterization of the mesoscale variability. A four-satellite configuration also improves the consistency between satellite altimetry and tide gauge data [9], and is critical for near real-time assimilation into numerical models, particularly important for operational oceanography and over coastal regions [8]. The main recommendation made by the international altimetry team is to improve the spatial and temporal samplings simultaneously, since smaller features and processes generally have a faster evolution [1].

Measurements provided by the MAGAL constellation are expected to augment those from the current operational satellite radar altimeters. When used in synergy with the latter, e.g. Sentinel-3A/B, Sentinel-6 Michael Freilich and Jason-3, MAGAL should contribute to accomplishing the desired four (or more) satellite configuration for improving the multi-mission altimetric dataset over open and coastal ocean. Sampling the ocean with a 5-day repetition cycle, between latitudes ±82º, and with ground tracks separated at the Equator by ~90 km, MAGAL is expected to increase the current observation capabilities of satellite altimetry.

### 2.2 Eddy Detection and Tracking

Eddies, in particular mesoscale eddies, are extremely important for energy, salt, and biogeochemical material exchanges and ocean production [10]. Efficient eddy detection and tracking is therefore crucial for advancing our understanding of ocean dynamics and circulation on the surface, at depth and in the deep ocean. Nonetheless, eddy detection and tracking are still two challenging problems, especially in areas of multiple eddies or areas rich in sub-mesoscale eddies. Present limitations for eddy detection and tracking using satellite altimetry are correlated to the temporal and spatial sampling provided by current satellite altimetry missions, since eddies can quickly change their status over several days. Moreover, observations of long-term time series within eddies are required. Therefore, it is imperative that satellite altimetry can provide data at smaller and faster scales, while extending their coverage to coastal areas and to high latitude regions.

MAGAL data, combined with *in situ* data (e.g. gliders, drones and oceanographic data surveys) are expected to allow eddy differentiation (e.g. regular anticyclones from mode-water eddies or regular cyclones from cyclonic thinnies), and to study their physical and biogeochemical dynamics.

### 2.3 Marine Debris Monitoring

Marine litter is one of the most prominent problems in the global ocean. However, the number of observations and studies on this phenomenon remains scarce and somewhat limited to specific locations in terrestrial and marine biomes, such as coastal areas [11]. One of the possible actions that can be implemented to develop effective policies to detect and monitor this phenomenon is satellite remote sensing (RS). Through satellite altimetry, provided data can eventually cover enough spatial and temporal sampling for litter analyses. Still, numerical modelling has been used globally to describe the movements of small plastic particles, unravelling spatial processes at different scales, i.e., from oceanic gyres to mesoscale eddies [11]. Results show, however, discordances when compared to those from *in situ* observations [12].

The possibility of improving present temporal and spatial altimeter sampling through the proposed MAGAL constellation, shall make it possible to detect and track in time more efficiently oceanic regions of accumulation of marine debris (e.g. frontal regions, mesoscale cyclonic and anticyclonic eddies, surface topographic effects). Present Lagrangian numerical models can be refined with higher resolution altimeter data, for example, to better detect and track mesoscale processes that contribute to the dispersion and/or sink of these particles along the surface and within the water column, towards the deep ocean.

### 2.4 *Inland Water Levels*

Satellite altimetry has been used for hydrological studies to monitor continental water surfaces over large inland water bodies such as lakes, reservoirs, rivers, estuaries, and flood plains, supporting new applications for hydrological studies. For many remote areas, satellite altimetry is the only available source of information [13], which in turn, can be used to calibrate and evaluate hydrological models [1]. Currently, the use of satellite altimetry for inland water studies is limited by the revisit time of past and current missions and by the size of the altimeter footprint, therefore allowing to monitor water level variation over a limited number of lakes and rivers [1].

The most significant limitations of current altimeters in water/land interface regions are, namely: 1) the large altimeter footprint (from 2.5 to more than 10 km diameter for Low Resolution Mode (LRM) measurements), and depending on surface roughness, improved to ~300 m in the along-track direction on Synthetic Aperture Radar (SAR) altimeters; 2) the contamination of the altimeter echoes by the existence non-water surfaces within the footprint, requiring tuned retracking for their correct identification; 3) difficulties in the computation of the necessary range and geophysical corrections; 4) the need for high-rate altimetry measurements to increase along-track data density; and 5) the geoid errors (reference datum over inland water bodies).

One of the major drawbacks of altimetric height measurements for water stage monitoring is the temporal sampling rate, where water levels over inland water bodies have traditionally relied on data from the reference missions having a 10-day repetition cycle (e.g. [14, 15]).

The MAGAL constellation can support the continuity of existing altimetry data services and improve their accuracy, resolution, and error characterization, leading ultimately to better understanding of the long-term evolution of inland water level due to climate change and water resources uses.

## 3 CONSTELLATION DEVELOPMENT

### 3.1 *Mission and Orbit*

The main goal of the MAGAL constellation is to provide scientific data with high temporal and spatial sampling time. Based on the defined use cases (Section 2), and respective needs, all satellites must be placed in Exact Repeat Orbits, flying over the same region at least every five days. The distance between ground tracks at the equator should not exceed 100 km. Additionally, the satellites must be placed in a Sun-Synchronous Orbit (SSO), which means that their inclination is about 98°. Due to power budget limitations, the altitude of the constellation should not exceed 600 km, otherwise the altimeter's power consumption would be too large. Since atmospheric drag might greatly affect the constellation, a minimum altitude is preliminarily set to 500 km.

Taking these restrictions into account, the number of satellites required for the mission was calculated according to the desired spatial sampling for a repetition time lower or equal to 5 days (Figure 1).

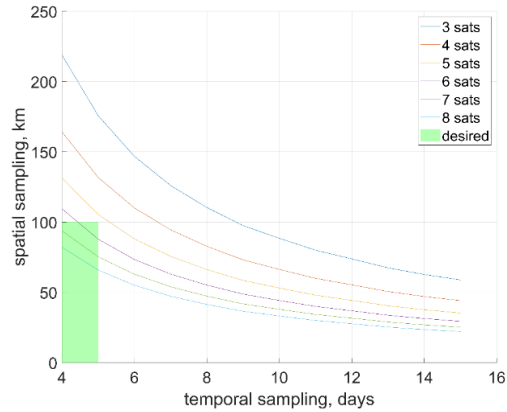

Figure 1. Number of satellites required and provided sampling. The green box represents the minimum requirement, with the curves for 6 or more satellites fulfilling it

Based on the analysis translated to Figure 1, at least six satellites are required to fulfil the mission needs. To establish the constellation altitude, some calculations were performed using averaged orbital elements. Despite the difference between the approximation and the real orbit altitude could be several kilometres, for preliminary orbit design they are sufficient and satisfactory. As the orbits are supposed to be circular, the eccentricity is set to zero. Equations for secular drifts due to the Earth's oblateness perturbation are, according to [16]:

$$\dot{\Omega} = -\frac{3nR_E^2 J_2}{2p^2} \cos i, \quad (1)$$

$$\dot{\omega} = \frac{3nR_E^2 J_2}{4p^2}(4 - 5\sin^2 i), \quad (2)$$

$$\dot{M} = n - \frac{3nR_E^2 J_2}{4p^2}\sqrt{1-e^2}(3\sin^2 i - 2). \quad (3)$$

Here $\dot{\Omega}$ is the Right Ascension of the Ascending Node (RAAN), $\omega$ is the Argument Of Perigee (AOP), $M$ is the mean anomaly, $n$ is the mean motion, $\mu$ is the Earth gravitational parameter, $a$ is the orbit semi-major axis, $R_E = 6\,378.14$ km is the Earth's radius, $J_2 = 1\,082.62 \times 10^{-6}$ is the dynamical form factor, $p$ is the orbit semi parameter, $e$ is the eccentricity, $i$ is the inclination. Since we consider SSO options only, the RAAN precession is set to be:

$$\dot{\Omega} = \frac{2\pi}{365.25} \frac{rad}{day}, \quad (4)$$

hence inclination and orbit semi parameter cannot be chosen independently. These orbits are supposed to be circular, so $e = 0$, $p = a$.

To provide periodic ground tracks, it is necessary to ensure the "orbit resonance", namely:

$$T_\Omega \cdot n_{days} = T_u \cdot n_{rev}, \quad (5)$$

where $T_\Omega = 2\pi/(\omega_E - \dot{\Omega})$, $\omega_E$ is the Earth's angular velocity, $T_u = 2\pi/(\dot{M} + \dot{\omega})$, $n_{days}$ is the repetition cycle in days, and $n_{rev}$ is the number of revolutions necessary to repeat the ground tracks. Although the final orbit selection will require higher fidelity models in trajectory design, this work presents the preliminary analysis for initial constellation design.

Combining Eqs 2-5 result in an estimate of the required orbit altitude for the desired revisit time (shown in Figure 2).

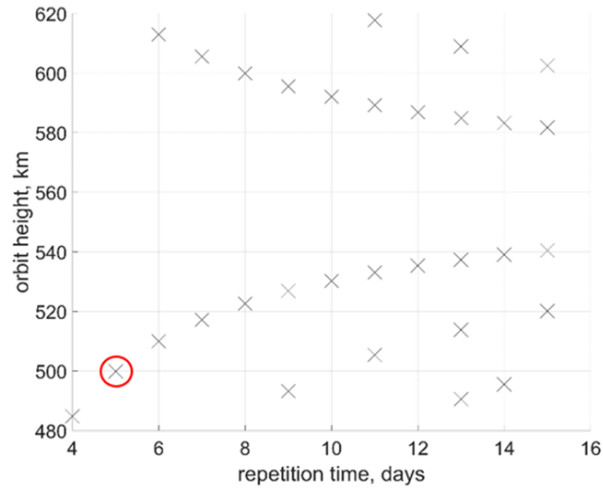

Figure 2. Dependence between the orbit height and the period of ground track repetition.

As can be seen in Figure 2, it is necessary to place the satellites at an altitude of 500 km to provide the intended 5-day periodic ground tracks. On the other hand, it is possible to consider other scenarios with 10 or 15-day periodic orbits, by placing two or three satellites, respectively, on the same ground track. Such configurations will still provide a 5-day repetition cycle of the constellation and almost the same spatial resolution. Additionally, this latter solution would allow for selecting higher orbits of 580 km up to 590 km to reduce the effect of atmospheric drag.

Figure 3 presents the ground tracks for the 5-day repetition cycle, at the left. For this simulation, the orbit altitude is 499.85 km, with an inclination of 97.4, the number of revolutions necessary to repeat the ground tracks is 76, leading to a spacing between adjacent ground-tracks of 87.88 km, at the Equator. A similar result is obtained for a 10-day repetition cycle (Figure 3 on the right side). Orbit altitude is 592.07 km, inclination is 97.75 degrees, the number of revolutions to repeat the ground tracks is 149, and distance between adjacent ground tracks is 89.65 km.

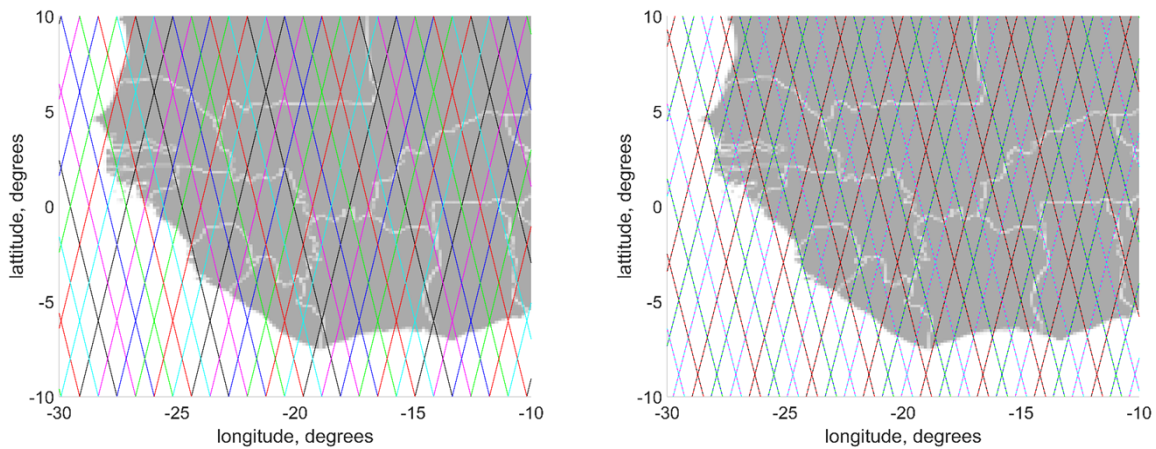

Figure 3. Ground tracks of the constellation for a 5-day repetition cycle configuration (left) and for a 10-day repetition cycle configuration, with two satellites at each ground track (right).

## 3.2 Altimeter Payload

One of the most used sensors in Earth observation missions, specifically in topographical and/or oceanographic studies, is the altimeter, measuring the distance between the satellite and the surface. One type of altimeter is the Radar Altimeter (RA), which can make range measurements with resolutions in the order of tens of centimetres [17], and is widely used in several missions, such as the Cryosat-2 [18] and Sentinel-3 [19].

RAs typically use SAR technology [20], which consists of transmitting a broad set of short-duration (nanosecond) electromagnetic wave pulses and measuring the time these pulses take along the way from satellite to Earth and back to satellite. Handling this data requires high-performance digital systems for time-of-flight measurement and signal processing, resulting in high energy consumption. Due to the size and mass limitations of a small satellite, the power available on the platform is very limited, making it difficult to implement a SAR system.

To overcome this difficulty, MAGAL will focus on developing a RA for oceanographic studies, including coastal areas, in a low-Earth orbit, compact and with low power consumption. Radar techniques such as Frequency Modulated Continuous Wave (FMCW) [21] guarantee less computationally demanding systems and consequently low power. FMCW radar altimeters are widely used in aeronautical applications, for this reason, and they are designed for low altitudes (500 m to 5 km) [22]. For space applications, the RA must be able to operate at much higher altitudes (200 km to 1 000 km), and a redesign of the system is necessary.

### 3.2.1 Architecture

As discussed above, the power available on a small satellite is very restricted, due to size and mass limitations. Thus, using technologies that require complex processing and algorithms, such as pulsed RADAR or SAR systems, is not feasible with current technology. A better solution is to use an FMCW RADAR system that guarantees computationally fewer demanding systems. The difference in efficiency between the two types of systems is due to the greater bandwidth and more complex algorithms to measure the time of flight of the signal required by pulsed RADAR and SAR, whereas the FMCW RADAR only processing is measuring the difference of frequencies and, at the same time, the signals usually have a lower bandwidth.

### 3.2.2 Operations Basics

The system will operate in the Ka-band, more specifically at 13 GHz, as this slot of frequencies is usually assigned for this type of operation for the current altimetry satellites in orbit. The maximum power consumption goal for the entire system is 20 W. Given that the power amplifiers are responsible for most of the consumption, the transmitted signal power ($P_{tx}$) will be about 1 W (30 dBm). From the transmitted power, it is possible to estimate the received signal power (reflected by the ground) using the RADAR equation [23]:

$$P_{rx} = P_{tx} \times \frac{\lambda^2}{(4\pi)^3 R^4} \times G^2 \times \sigma, \tag{6}$$

$$P_{rx}(dBm) = P_{tx}(dBm) + 10 \times \log_{10}\left(\frac{\lambda^2}{(4\pi)^3 R^4}\right) + 2 \times G(dB) + \sigma(dB), \tag{7}$$

where R is the distance from the RADAR to the target (in this case, the distance from the satellite to the Earth's surface, Figure 4), G is the antenna gain, and σ is the Radar Cross Section (RCS) value.

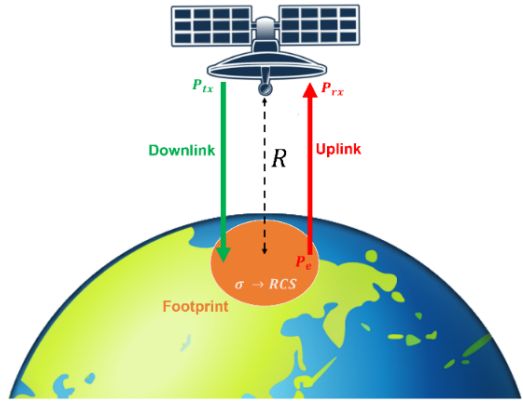

Figure 4. Representation of the propagation path of electromagnetic waves

### 3.2.3 Antenna

This RADAR altimeter will use a deployable dish antenna with a diameter of 1.5 meters. The antenna dimensions and design were selected to make the antenna feasible for the MAGAL spacecraft, while maintaining a good antenna gain and footprint. In addition, this antenna will have a maximum envelope of 50 cm x 30 cm x 10 cm (when not deployed) and an expected mass of around 12.5 kg.

The selected antenna will provide an antenna gain of around 44 dBi and a footprint (coverage area on the surface of the Earth) of around 50 km$^2$. These values are essential to have an operational system because it defines the Radar Cross Section (RCS).

### 3.2.4 Radar Cross Section

The RCS is a measure of the ability of a given target to reflect the signal and depends on the area of the section exposed to the RADAR signal, reflectivity, and directivity. In the literature, it is usual for the RCS value to be normalized to the measurement area (A). This normalized RCS value ($\sigma^0 = \frac{\sigma}{A}$) can also be called the radar backscatter coefficient [24].

However, the value needed for the Radar equation calculations is σ, which is calculated by multiplying $\sigma^0$ and the coverage area. It is known that the beamwidth of the antenna imposes the maximum value of the coverage, but not all this area will contribute to the RCS.

Despite the RCS or the radar backscatter coefficient, radar systems are typically analysed by assuming that the target is in the far-field relative to the radar so that an incident plane wave can be considered [25]. This approach reduces the analysis's complexity and simplifies the RCS calculation, which can be considered constant for any distance to the target. Therefore, most RADAR literature considers that the target is always in far-field conditions [26]. However, in many RADAR applications, especially in the RA cases (Figure 5), where the distance and area of the target are very high, the far-field assumption cannot be accepted [26].

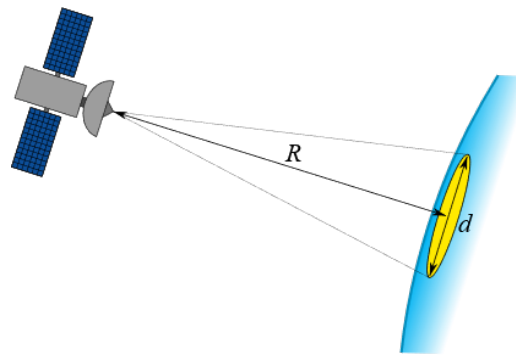

Figure 5. Antenna observation area for RA applications

Considering the scenario presented in Figure 5, one can consider that the boundary between the far-field and near-field is given by [27]:

$$R > \frac{2d^2}{\lambda}, \tag{8}$$

where $R$ is the distance from the satellite to the target, and $d$ is the diameter of the observation area. For this case, the limit value of $d$ would be around 75.8 meters.

Using statistical models and simulations, analysing the RCS behaviour for the near-field and far-field cases [28] (Figure 6), it is possible to conclude that the best situation would be in the boundary condition of the transition between the two fields.

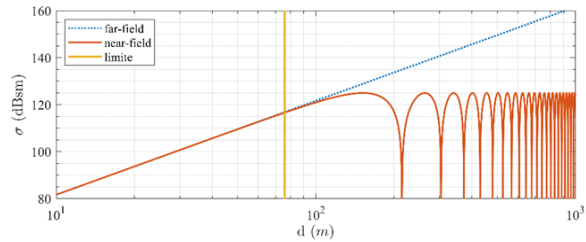

Figure 6. RCS as a function of observation diameter for a satellite height of 500 km and a signal frequency of 13 GHz

### 3.2.5 Overall System

Once all these factors have been calculated, the RA radio system can be sized, making sure that the received signal has enough power and signal-to-noise ratio to be received and obtain valuable data.

As a control unit, a Field Programmable Gate Array (FPGA) will process the received signal, obtain the range profile (Fourier transform of the frequency differences), calculate the distance to the target and the Received Signal Strength Indicator (RSSI) of the *Rx* power.

Expected accuracy of a single distance measurement, prior to its correction and validation, shall be better than 20 cm at a maximum rate of 20 Hz. With range profile data and oversampling w.r.t. the minimum required sampling data (1 Hz), it is expected to improve distance accuracy to 2-3cm.

In addition, the RA system, excluding the antenna, will be a very compact system (15 cm x 10 cm x 10 cm) and light (below 3 kg). It will also have internal memory capability to store the range profile data at 1 Hz for a complete orbit before downlinking it to the ground station.

### 3.3 Spacecraft Platform

In the case of a small satellite with a radar altimeter, satisfying power and size requirements are the main engineering and design challenges, and although there are small satellite mission concepts, using a small constellation (two to six 6U CubeSats) and a radar altimeter with 1 W [29], most missions resort to conventional satellites (e.g. the SARAL spacecraft with approximately 400 kg [30] or the 550 kg of Jason 3 [31]).

Within MAGAL, an innovative satellite architecture, capable of overcoming some of the limitations imposed by traditional CubeSats, is being designed. To have a cost-effective constellation, i.e. at the lowest cost possible while assuring the needed quality and reliability, the main bus systems have been split into those which can be acquired as Commercial Off the Shelf (COTS) and those that need to be developed, depending both on the system's criticality and know-how among MAGAL partners.

The intended platform is targeted to have a total mass below 100 kg, with a form factor around 1 m x 0.4 m x 0.4 m (more compact than the current smaller conventional altimetry mission, SARAL). It will be capable of carrying the antennas for the altimeter, as described in Section 3.2, and a set of deployable solar panels to generate sufficient power for all mission phases.

A particular system under development in MAGAL is the propulsion one. As spacecraft tend to be more compact and smaller, so does their propulsion system. Although proper dimensioning of these systems highly depends on mission delta-v requirements, scaling down a propulsion system has its limits, due to both inherent system limitations as well as its integration with the remaining systems.

Propulsion systems provide thrust and torques for spacecraft attitude and orbit control. The main manoeuvrings are orbit transfer, insertion, maintenance, deorbiting, and attitude control. For these purposes, different propulsion systems are used, rooted in different architecture and technologies, namely chemical, electric, and propellant-less ones, in which the latter have been used solely for small-scale demonstrations.

Electric propulsion systems deliver an order of magnitude greater total impulse than chemical systems. However, due to the low thrust-to-power ratio, electric systems demand high power supply to provide low thrust values. Furthermore, electrical power systems for small satellites are difficult to implement, particularly when it comes to generation, storage, and distribution of electrical power for all equipment and payloads in a constrained volume, constantly requiring investments to increase the power-to-mass ratio. Therefore, despite the high total impulse, electric systems may need to operate hundreds or thousands of hours compared to the seconds or minutes that chemical systems demand to provide the same impulse.

Chemical propulsion can provide higher thrust-to-power ratios and rapid manoeuvres for both attitude and orbit control and are, therefore, the most natural option for small satellites which require orbit manoeuvrings, when the total impulse budget is enough to meet the mission requirements.

Chemical propulsion systems have extensive flight heritage from larger satellites, whereas for small spacecraft there are few available miniaturized components, with most of these under development to meet this relatively new space market scenario. Some equipment are adaptations of existing and flight-proven components from large satellites propulsion systems, yet many are brand-new developments to fulfil the small spacecraft trend.

Cold gas systems, the simpler of the chemical solutions, are the most flight-demonstrated type for small satellite missions, despite their significatively lower specific impulse. Conversely, mono/bipropellant types are one of the most mature technologies. This impacts on delivered delta-V, yielding that chemical system type decision depends on the mission and propulsion requirements.

Mono/bipropellant systems require less volume than cold gas systems to comply with the same mission. In terms of orbit manoeuvrings, the cold gas systems are out of scope, applicable for only attitude control. Targeting for the best possible performance and more flexibility in the kind and number of orbital manoeuvres, mono/bipropellant propulsion system would be a natural choice, resulting however in an increased cost, complexity and ultimately, risk to the project.

To support the decision for MAGAL, major pros and cons for each chemical propulsion system have been listed in a trade-off, namely considering the associated manoeuvres, lifetime, and in-flight operations to fulfil the MAGAL needs. Table 1 presents a qualitative summary of this trade-off for each propulsion system. Based on these results, the mono/bipropellant has been selected as the most promising solution, and the one to be pursued.

Table 1. Pros and cons of different propulsion system types (++ - "most advantageous"; + - "Advantageous"; - - "Disadvantageous")

| Technology | Specific impulse | Mass/Impulse | Volume/Impulse | Power/Impulse | Complexity | Flexibility | Cost |
|---|---|---|---|---|---|---|---|
| Cold gas | - | + | + | ++ | ++ | - | ++ |
| Mono/Bipropelent | + | ++ | ++ | + | + | ++ | + |
| Electric | ++ | - | - | - | - | + | - |

## 3.4 Data Centre

### 3.4.1 Basic Infrastructure

For the next years, satellite data is expected to increase. According to projections, from 2019 to 2033, 20425 satellites are expected to be launched, having North America leading the way, followed by European companies [32]. Hence, one of the major topics is how to make the data available and sustainable to both research and commercial users. Although there are some commercial Data Centres, it is important to develop a differentiating Data Centre, capable to have additional services besides what is already on the market for specific market niches.

During the development of the MAGAL project, a DAC is going to be developed. Once the satellite is launched into orbit, data will be acquired from sensors and downlinked to the ground. After the data is downlinked, it will enter the DAC to be displayed to end-users. Therefore, the DAC is a web platform that is both a repository for large volumes and also a platform capable to deliver and disseminate services directly to customers and clients (Figure 7).

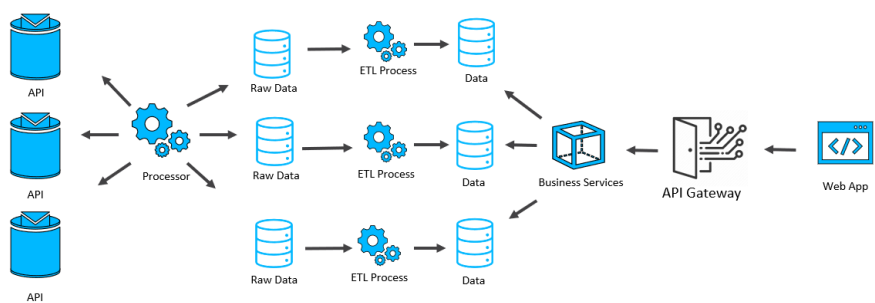

Figure 7. Data Analysis Centre architecture

The architecture of the DAC follows a typical software web development, having a back end (data acquisition), an ETL (Extract, Transform and Load the data), and a front-end (dashboard) development as can be seen in Figure 7. MAGAL DAC is being developed with a modularized concept, using a micro-services architecture, ensuring both horizontal and vertical scalability, allowing multiple web interfaces, and guaranteeing a design security approach.

for the ETL, the process consists in the development of all the phases for data treatment. The ETL follows a two-step approach for data extraction and data parallelization (data ingestion). The data extraction allows to obtain data from a database (software-as-a-service, or SaaS), to be further replicated to a destination. Data processing parallelization, on the other hand, is a computational method for running two or more Central Processing Units (CPUs), capable to handle separate parts of an overall task. This methodology allows to make use of multi process tools.

After back-end and ETL development, a front-end (dashboard) will be developed. As mentioned above, using a microservices approach allows to encapsulate each business capability into individual services and lets them interact with each other. Compared with typical software monolithic development, this methodology allows a software mindset, reducing development complexity (maintaining and updating becomes easier), where developers can take full responsibility ("you built it, you run it, you own it" mindset), having increased agility (each microservice works independently) and resilience (fewer points of failure), and more effective scaling (each microservice is small and fully functional on their own) and specialization (each microservice uses the best-suited technology for the task in hand). The front-end also considers the development of the Application Programming Interface (API) gateway, built upon a multi-tenancy approach (a single instance of a software application serves multiple customers).

The dashboard (web app) allows scalability for different purposes of multi-tenancy actors, following a modular approach with design system concepts like atomic design (five steps: atoms, molecules, organisms, templates, pages). Following a modular approach, each tenant can have his User Interface and some specific features required by its actors, always following a consistent design.

From an infrastructure point of view, the system will be developed considering a distributed system to be able to be managed independently, adapting the microservices architecture with the use of Cloud services. It will also be capable to integrate data form the MAGAL constellation, as well as from other public and private sources.

### 3.4.2 Data Products

For the MAGAL project, the platform will acquire the data in different development status (from L0 to L4). In terms of DAC products, the first product data, the L0, represents data at an initial development status (only has some minor modifications, e.g. sensor calibration).

Along-track (Level-2) and gridded altimeter product types will also be generated. These products form the basis for the use cases defined in Section 2. For the along-track products, the main customers are science, aquaculture and offshore, oil and gas, and the wind energy industries. In terms of timeliness, Near Real Time (NRT), Short-Time Critical (STC) and Non Time Critical (NTC) products will be generated and delivered to users. NRT products, delivered within 3 hours after data acquisition (one file per pass, i.e. pole to pole data), shall be used for marine meteorology, ocean-atmosphere studies, and operational oceanography. STC products, delivered within 48 hours after data acquisition, correspond to one file per pass and should contain some consolidated data (e.g. preliminary restituted orbit) and will be disseminated mainly for oceanography studies and operations, as well as ocean forecasting. NTC products will be delivered with a delay of one month after data acquisition to allow for the consolidation of the orbit and the range and geophysical corrections. Again, one file per pass and/or one file per cycle will be available. This product type is mainly used for sea level, geophysical and ocean circulation studies, operational oceanography, assimilation into numerical models, biological and climate studies, and bathymetry.

Level-2 along-track products shall contain cycle, pass, time, latitude, longitude, altimeter range (corrected from instrumental errors) and its standard deviation, sea level anomaly (SLA) and the number of valid measurements used in SLA computation, significant wave height, backscatter coefficient ($\sigma_0$), range and geophysical corrections, information on data quality, and auxiliary data (e.g. mean sea surface, geoid). Along-track data will be provided at 1 Hz over ocean, and at 20 Hz, at least, over coastal, and inland water regions. In-house algorithms, developed by the team, to derive the range and geophysical corrections will be used.

For inland water studies, only STC and NTC products will be generated and disseminated to users mainly meant for reservoir/river level studies, flood management and forecasting, reservoir operations, and calibration of river/lake models. Level-2 products will have the same content as those for ocean applications, except that some range and geophysical corrections may not apply. Main customers are coastal authorities (governmental and public) and port authorities.

Gridded products will be generated for ocean circulation, ocean bathymetry and mesoscale studies, operational oceanography and assimilation into numerical models, eddy analysis, and biological and climate studies. These include SLA gridded products, one file per cycle, with a quarter degree spatial resolution and generated from Level-2 along-track data (NRT/STC/NTC) using objective analysis. Absolute dynamic topography (ADT) grids will be generated at the same SLA spatial and temporal resolutions, using a state-of-the-art mean dynamic topography available through Satellite Altimetry data providers. Zonal and meridional components of geostrophic currents, as well as geostrophic velocity anomalies, will also be generated (one file per cycle, same spatial and temporal resolutions of SLA and ADT gridded products). Geostrophic currents will be generated by means of a finite difference scheme using a two-dimensional nine-point computational stencil, for latitudes outside the 3°S/3°N band. In the equatorial region, the β plane formulation shall be used. Eddy Kinetic Energy (EKE) grids will be calculated from zonal and meridional geostrophic velocity anomalies.

For eddy detection and tracking and marine debris case studies, besides these along-track (NRT/STC/NTC) and gridded products derived from MAGAL, ocean colour bi-optical parameters and Near-InfraRed (NIR) Sea Surface Temperature (SST) products (also NRT/STC/NTC), both, at least, with 1 km resolution, and *in situ* surface and water column oceanographic data are needed. Therefore, the DAC will also disseminate these products, which shall allow a better understanding of the: ocean dynamics, mesoscale, and sub-mesoscale circulation; eddies contribution to the mean ocean circulation; biogeochemical processes in the ocean; eddy detection, tracking, type, and life-history (evolution); contribution of mesoscale eddies to North Atlantic circulation; climate studies. Maps identifying eddies type (cyclonic, anti-cyclonic, mode-water, thinning), eddies frontal regions, convergence, and divergence regions important for biology, waves, and wind fields for aquaculture sites, among others, will also be provided. Main users of these products are customers in the ocean sciences, fisheries services, official authorities, public authorities, and aquaculture.

## 4 SUMMARY

The MAGAL Constellation is being designed as an elegant, compact, and powerful solution with which to fill current knowledge gaps in global ocean circulation variability at regional and local levels. MAGAL addresses thus four main use cases, driven by a scientific need for higher spatiotemporal resolution on ocean circulation data: (1) characterization of the mesoscale variability at local and regional scales to support operational oceanography; (2) eddy detection and tracking; (3) monitoring of marine debris pathways; and (4) the monitoring of inland water bodies' level.

To achieve these use cases, a small satellite platform with a small, low cost, optimised altimeter is being designed. This altimeter will operate at a frequency of 13 GHz, with a Frequency Modulated Continuous Wave (FMCW) architecture, selected for its lower power consumption (20 W consumption for 1 W transmission RF power) than traditional pulsed radar altimeters. The signal received by the payload will then be treated using digital signal processing and a Field Programmable Gate Arrays (FPGA) approach. To guarantee an adequate observation footprint and RF link budget, for the intended goals, the payload antenna will be a deployable dish, with up-to 1.5 m diameter.

A configuration no larger than a 24U small satellite will allow the integration of all the needed systems. To reduce launch costs of the constellation, all six satellites will be launched simultaneously, in the same orbital plane, at the same time and vehicle.

Beyond flying in a SSO orbit, which allows the MAGAL constellation to have a worldwide coverage, its main added value and improvement with regard to the current radar altimetry constellations is its temporal sampling. MAGAL constellation is proposed to have a repeat ground track period of 5 days, which is a significant improvement to the current solutions, keeping similar spatial samplings (about 100 km at Equator, with a global ground track mesh that is homogeneous in the longitudinal direction and providing the same spatiotemporal sampling for symmetric latitudes) and distance accuracy. This improvement of the temporal sampling can lead to a significant improvement of many services and products.

Currently, besides the use cases, the mission requirements have been set and a baseline of the system and equipment established, where key accuracy parameters for MAGAL constellation are fully in line with current altimetry. These allow the kick-off of the development of both the platform, including its payload, and of the DAC and corresponding algorithms. These will be assessed during the MAGAL project, which will end with the Critical Design Review and preparation for implementation.

## ACKNOWLEDGMENTS

The MAGAL Constellation project (Nr. 033688) is co-financed by the ERDF - European Regional Development Fund through the Competitiveness and Internationalisation - COMPETE 2020, LISBOA 2020, PO ACORES 2020 (ACORES-01-0145-FEDER-000129) and by the Portuguese Foundation for Science and Technology - FCT under the UT Austin-Portugal International Partnership programme.